\documentclass[aps,prl,showpacs,twocolumn]{revtex4}

\usepackage{graphicx}
\usepackage{longtable}
\usepackage[ansinew]{inputenc}

\begin{document}

  \bibliographystyle{prsty}

  \title{Molecular Orbital Shift of Perylenetetracarboxylic-Dianhydride}
  \author{J. Kröger}\email{kroeger@physik.uni-kiel.de}
  \author{H. Jensen}
  \author{R. Berndt}
  \affiliation{Institut für Experimentelle und Angewandte Physik,
    Christian-Albrechts-Universität zu Kiel, D-24098 Kiel, Germany}
  \author{R. Rurali}
  \author{N. Lorente}
  \affiliation{Laboratoire Collisions, Agrégats, Réactivité,
    UMR 5589,
    Université Paul Sabatier, F-31062 Toulouse, France}

  \begin{abstract}
Using low-temperature scanning tunneling microscopy we find that
perylenetetracarboxylic-dianhydride on Au(788) exhibits three coexisting
adsorption phases. Single-molecule tunneling spectroscopy reveals orbital
energies, which differ in the different adsorption phases. Density functional
theory calculations associate the experimentally observed submolecular
corrugation to the spatial distribution of the second-to-lowest unoccupied
molecular orbital. We tentatively attribute the orbital shifts to a
varying number of hydrogen bonds.

  \end{abstract}

  \pacs{68.37.Ef,73.20.At,73.61.Ph}

  \maketitle

Interest in organic molecular films is fueled by
technological advances in organic devices \cite{dim99,oij99}, potential
applications in molecular switches \cite{che99,lah03}, and
functionalization of surfaces \cite{tan87,bur93,tay97} as well as by
the fundamental understanding of the
film-metal system.
Typically, the geometric and electronic structure
of the molecule-substrate interface is of substantial importance.
Therefore considerable effort has been made to investigate this interface
and to maximize its homogeneity \cite{for97}. Owing to the complexity of
molecular shapes and interactions, however, perfectly homogeneous
interfaces usually remain elusive and coexistence of structural domains
has been reported for various molecular systems \cite{wit04}. Detailed
characterization of the variability of the electronic structure on a
single molecule scale requires high lateral resolution and has,
consequently, hardly been achieved \cite{lud94,toe01}.

A model system of molecular electronic studies at the molecular level is
$3,4,9,10$-perylenetetracarboxylic-dianhydride (PTCDA). The adsorption
geometry of PTCDA on metal surfaces was analyzed previously as excellently
reviewed by Witte and W\"oll \cite{wit04} and Barlow and Raval \cite{smb03}.
Adsorption characteristics on
Au(111) deserve special interest since this surface exhibits a
$22\times\sqrt{3}$ reconstruction. Chizhov {\it et al.} \cite{chi00}
found that room-temperature adsorption leads to two coexisting adsorbate
geometries, namely a herringbone structure and a square
phase, where four PTCDA molecules adsorb on the boundary of a square with
equivalent axes of adjacent molecules including an angle of $90^{\circ}$.

\begin{figure}
  \includegraphics[bbllx=10,bblly=200,bburx=580,bbury=640,width=85mm,clip=]{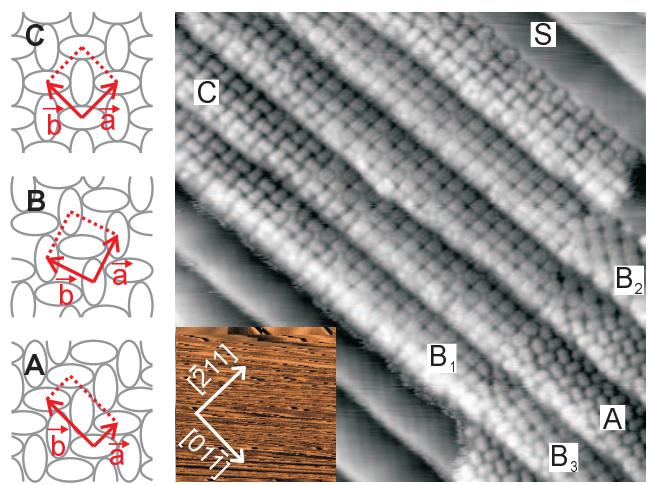}
  \caption[zoom]{Constant-current STM image of $0.25\,\text{ML}$ PTCDA on Au(788)
   (310 \AA\ $\times$ 310 \AA, $V = 0.75$ V, $I = 200$ pA).
   Crystallographic directions along and perpendicular to step edges are indicated.
   A, B$_n$, and C denote different PTCDA adsorption phases.
   Inset: clean surface area with v-shaped
  discommensuration lines.
  Left hand side: sketches of the observed PTCDA superstructures including lattice
  vectors.}
  \label{zoom}
\end{figure}

Here, we report on low-temperature scanning tunneling microscopy (STM) of
PTCDA on Au(788). We observe
three coexisting PTCDA superstructures on this surface. Single-molecule
tunneling spectroscopy reveals a peak above the Fermi level whose energy
strikingly varies between these geometric superstructures.
According to density functional theory (DFT)
calculations for the molecule-surface complex the experimentally observed
spectral feature is predominantly due to the second-to-lowest unoccupied
molecular orbital (LUMO+1). This interpretation is in some contrast to
previous work for PTCDA on other metal surfaces where experimental
features have been attributed to the LUMO on the basis of calculations for
the free molecule \cite{lud94,glo98,mer04}. We argue that the observed
shift of the orbital energy is most likely caused by hydrogen bonds.

\begin{figure*}
  \includegraphics[bbllx=10,bblly=325,bburx=580,bbury=515,width=180mm,clip=]{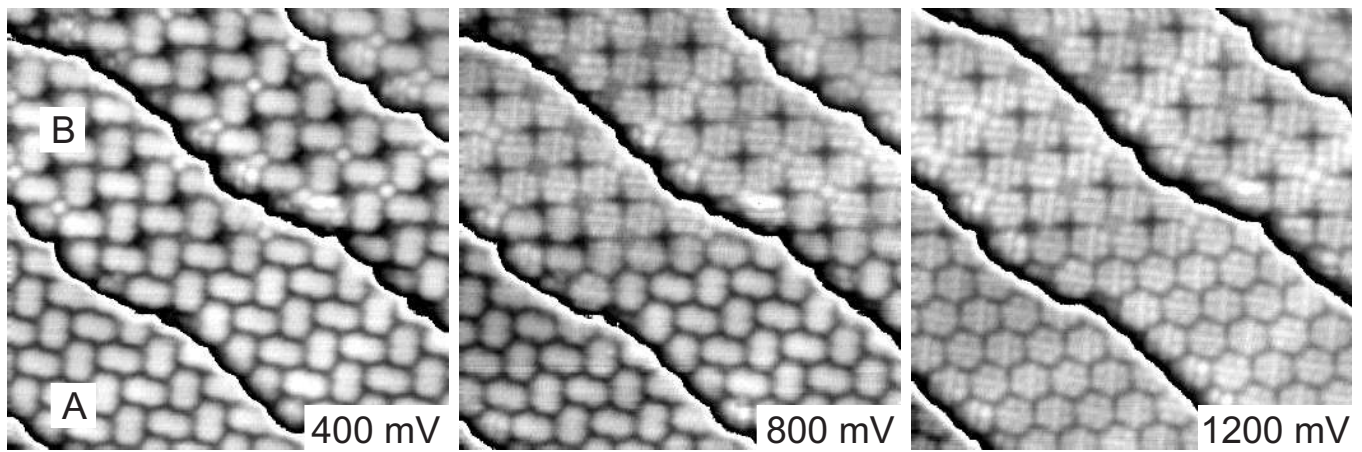}
  \caption[gal]{Sequence of constant-current images recorded at
  characteristic sample voltages.  The top (bottom) part of each image shows B-domain
  (A-domain) molecules ($120\,{\rm\AA}\times 120\,{\rm\AA}$,
  $I = 200\,{\rm pA}$).  A periodic grayscale is used to highlight the contrasts
  on each terrace.}
  \label{gal}
\end{figure*}

We used a home-built ultrahigh-vacuum (UHV) low-temperature STM with a
base pressure of better than $1\times 10^{-8}\,{\rm Pa}$. The Au(788)
sample and W tips were cleaned by argon ion bombardment and annealing.
Crystalline order was checked by low-energy electron diffraction, and
surface cleanliness was monitored by STM\@. PTCDA was evaporated onto the
Au(788) surface held at room temperature from a degassed tantalum
crucible keeping the pressure of the UHV recipient below $1\times
10^{-7}\,{\rm Pa}$. The rate of deposition, monitored by a quartz
microbalance, was $0.5\,{\rm ML}\,{\rm min}^{-1}$. One monolayer (ML) of
PTCDA is defined here as one molecule per $100\,{\rm\AA}^2$
(corresponding to the density of the herringbone superstructure). The
prepared sample was then cooled down to $4\,{\rm K}$ within $30\,{\rm
min}$. For detection of the differential conductivity (${\rm d}I/{\rm d}V$)
we applied a standard lock-in technique. The tip state was checked by spectra
of the electronic surface state of clean Au(788) acquired before and after
measuring the ${\rm d}I/{\rm d}V$ signal of single PTCDA molecules.

Density-functional calculations were performed with the code
{\sc Siesta} \cite{siest} using a double-$\zeta$ basis set plus
polarization functions \cite{ang02} for the valence electrons and the
Generalized Gradients Approximation (GGA) \cite{per96} for the
exchange-correlation functional. The Au(111) surface
serves as a model of the terraces of Au(788). Various supercell sizes in
a slab geometry were explored to take into account different packing
densities of PTCDA superstructures. For simulating experimental spectra of
the differential conductivity we calculated the local density of states
at the tip apex being $7\,{\rm\AA}$ above the surface. Due to the weak
molecule-surface
interaction local and semi-local approaches of DFT fail to yield the
correct adsorption parameters of PTCDA on Au(111). Here, we are
interested in a qualitative analysis of the electronic structure.
A recent incidence x-ray standing wave experiment performed on
Ag(111)-PTCDA determines the vertical distance of the perylene core
to $\approx 2.8\,{\rm\AA}$ \cite{aha05}.
Hence, we have frozen the height of the
molecular atoms at $2.8\,\text{\AA}$ and allowed a full
relaxation of the other molecular coordinates in order to find the
correct packing geometry for the two molecules constituing our supercell.
Our calculations show that increasing the
adsorption height by $1\,{\rm\AA}$ leads to a basically free molecule,
which slightly interacts with the surface. Decreasing the
molecule-surface distance by $0.3\,{\rm\AA}$ leads to a very
large mixing of molecular
orbitals because the molecule is already probing the strongly repulsive
substrate potential.

Figure \ref{zoom} shows a constant-current STM image of $0.25\,\text{ML}$
PTCDA on Au(788). Crystal directions along ascending steps,
$[\overline{2}11]$, and parallel to the step edges, $[01\overline{1}]$,
are indicated. The inset shows a small-scale STM image of a clean area to
highlight the discommensuration lines. These lines of higher corrugation
are stacking fault domains separating face-centered cubic (fcc) and
hexagonal close packed (hcp) surface layer stacking areas, similar to the
case of the $22\times\sqrt{3}$ reconstruction on Au(111). Au(788) is thus
patterned in two dimensions: terraces
with a $38\,{\rm\AA}$ periodicity along $[\overline{2}11]$ and
discommensuration lines with a periodicity of $72\,{\rm\AA}$ along
$[01\overline1]$. We find that the Au(788) reconstruction is not
perturbed by PTCDA adsorption. A similar observation has been reported
previously for the $22\times\sqrt{3}$ reconstruction on Au(111)
\cite{chi00}.

Three adsorption phases of PTCDA coexist on Au(788) (labelled
A, B$_n$, and C in Fig.\ \ref{zoom}). In domain A we find the herringbone
structure with lattice vector dimensions $a=(11.2\pm 0.7)\,{\rm\AA}$ and
$b=(17.9\pm 0.9)\,{\rm\AA}$ (sketches on the left hand side of
Fig.\ \ref{zoom}). Three prevalent orientations
of the shorter unit cell edge with respect to $[\overline{2}11]$ are
observed, namely $(33\pm 4)^\circ$, $(50\pm4)^\circ$, and $(77\pm
3)^\circ$. For Au(111), a different set of orientations ($22^\circ$,
$39^\circ$ and $55^\circ$) was reported \cite{sch99} and rationalized in
terms of a minimum misfit between the molecule film and the substrate. On
a stepped surface like Au(788) this effect appears to be suppressed by
the possibility to align the molecules with the step edges.

Superstructure B is a square phase with a lattice constant of
$a=b=(14.4\pm 1.2)\,{\rm\AA}$. Equivalent axes of
adjacent molecules are rotated by $90^{\circ}$ with respect to each
other. Several rotational domains, labelled B$_1$, B$_2$, and B$_3$  in
Fig.\ \ref{zoom}, are observed. Chizhov {\it et al.} reported a similar
superstructure on Au(111) \cite{chi00}.

Adsorption phase C differs from structure B by the end of a molecule
pointing towards the center of an adjacent molecule.

\begin{figure}
 \includegraphics[bbllx=0,bblly=0,bburx=450,bbury=300,width=85mm,clip=]{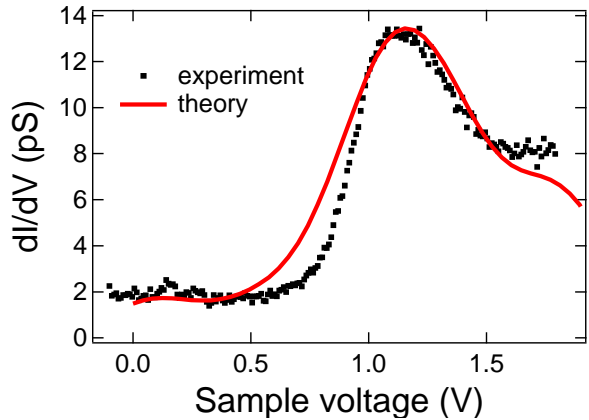}
 \caption[spec]{Experimental (crosses) and calculated (full line, fitted to the
  experimental maximum, the peak includes a numerical broadening of 0.25 eV)
  $\text{d}I/\text{d}V$ spectra of a single PTCDA molecule in the herringbone
  adsorption domain (A). Feedback loop parameters: $I = 200\,{\rm pA}$,
  $V = 1.4\,{\rm V}$.}
  \label{spec}
\end{figure}

We now turn to the electronic properties of the PTCDA molecules in the
superstructures.  Figure \ref{gal} presents STM images recorded at three
characteristic voltages. The images show four PTCDA covered Au(788)
terraces where A and B domains are visible in the lower and upper parts,
respectively. In B domains, circular protrusions are frequently observed
between four PTCDA molecules. We attribute these protrusions to single
trapped Au atoms in analogy to single Fe atoms surrounded by trimesic
acid molecules \cite{spi03}.  At $400\,{\rm mV}$ all PTCDA molecules are
imaged as structureless rectangles. For $800\,{\rm mV}$ topographic
changes of the A phase are hardly visible while intramolecular structure
emerges in the B phase, the former rectangular shape of the molecules
being lost in favor of a hexagonal shape. Finally, at $1200\,{\rm mV}$
submolecular structure is observed for both domains.  These observations
strikingly demonstrate that the electronic structure underlying the
intramolecular contrast depends on the adsorption phase of PTCDA
molecules.

To interpret the data we first trace the origin of submolecular structure
back to the involved molecular orbitals. Figure \ref{spec} shows the
differential conductivity ${\rm d}I/{\rm d}V$ as acquired in the center
of a single PTCDA molecule embedded in the herringbone adsorption domain.
We find a single peak centered at $\approx 1.1\,{\rm eV}$. The calculated
${\rm d}I/{\rm d}V$ spectrum, shown as a line, matches the experimental
spectrum. In the calculation, the main contribution to the conductivity
originates from the density of states of the LUMO+1.
The LUMO+2 is basically degenerated with the LUMO+1 and
also contributes to the spectroscopic
signature giving a certain broadening of the peak.
Since the intramolecular structure
for herringbone-phase molecules appears above $\approx 1\,{\rm V}$ we
ascribe most of the submolecular structure to the spatial distribution of
the LUMO+1\@.

\begin{figure}
\includegraphics[bbllx=0,bblly=0,bburx=450,bbury=300,width=85mm,clip=]{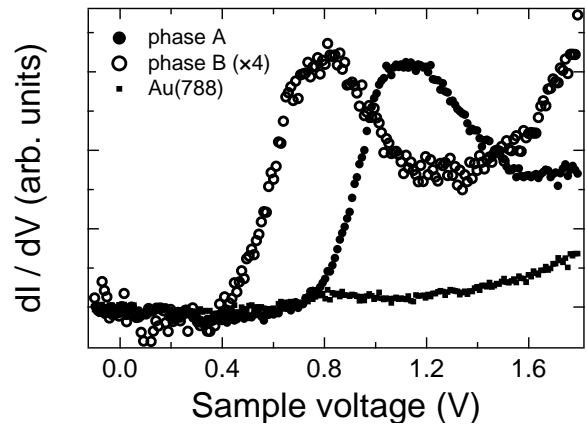}
\caption[shift]{Spectra of $\text{d}I/\text{d}V$ vs.\
  sample voltage on bare Au(788) (dots), on B-domain PTCDA (open circles),
  and on A-domain PTCDA (solid circles). Maxima of ${\rm d}I/{\rm d}V$
  occur at $\approx 0.8\,{\rm V}$ and $\approx 1.1\,{\rm V}$
  for B- and A-domain PTCDA, respectively. The feedback loop was
  frozen at $I = 200\,{\rm pA}$, $V = 850\,{\rm mV}$. Data for B-domain
  PTCDA was multiplied by 4.}
  \label{shift}
\end{figure}

We now return to the differences between domains A, B, and C observed in
Fig.\ \ref{gal}. Intramolecular structure for the square adsorption
domain B appears at lower voltages than for the herringbone phase.
Molecules at a domain boundary surrounded from either side by molecules
in different arrangements and likewise molecules in adsorption phase C
reveal submolecular structure for intermediate voltages.  Obviously, the
energetic position of the LUMO+1 depends on the local environment of the
individual molecule and thus differs for molecules from different
domains. We further tested this hypothesis by single-molecule
spectroscopy  of the differential conductivity acquired above the center
of B and A phase molecules (Fig.\ \ref{shift},
open and solid circles, respectively). The spectral signature of
B phase PTCDA is located at $\approx 0.8\,{\rm eV}$, {\em i.\,e.}
shifted by $\approx 0.35\,{\rm eV}$ compared to the A phase.
Owing to the similar intramolecular contrast in both phases
we can safely attribute this peak
to the LUMO+1\@. Since the feedback loop was frozen at $850\,{\rm mV}$ the
LUMO+1 contribution of the A-phase molecule has not yet developed leading to
a smaller tip-molecule distance than in the case of the B-phase molecule. To
show both peaks on the same scale we multiplied the data acquired on the
B-phase molecule by a factor of 4. Spectra of occupied electronic states,
{\it i.\,e.},
at negative voltages ($-2.5\,{\rm V}\leq V\leq 0$) turned out
to be featureless. Our calculations predict,
for instance, the LUMO of PTCDA to be pinned at the Fermi level~\cite{aha05}.
The
spatial distribution of the LUMO density of states is found to adhere strongly
to the molecule and to be less evanescent into vacuum than the LUMO+1.
This leads to its efficient suppression from
 the d$I$/d$V$ spectra (as observed in our experiments).

The calculations indicate that apart from the molecule-substrate
interaction the intermolecular one plays a role.  The latter is
mediated by hydrogen bonding between adjacent molecules which is poorly
accounted for by our numerical approach. Nevertheless, our calculation
for free standing molecules at the adsorbed geometries yields
a widening gap between molecular levels.
As a consequence the packing
density and the actual adsorption geometry influence the energetic
position of molecular orbitals.
The ability of DFT to reproduce workfunction changes even in
the case of physisorbed molecules \cite{mor04}, led us to compute and compare
the workfunction changes with the measured ones. The experimental
change is of the order of
0.1 eV which agrees with the change of the two molecular phases frozen at
2.8 \AA~from the surface atomic plane, implying that the molecules
lie at the same molecule-surface distance for both phases.
However the ${\rm d}I/{\rm d}V$ calculations show small differences between both
phases and are unable to explain the significant experimental orbital shift.
Different combinations of molecular heights for the two phases led
to inconsistent results and to disagreement with the measured workfunction
change. Hence, we conclude that the orbital shift is not due to
a different molecule-surface distance for the two phases.
Other effects such as the influence of the tip's electric field have
been included in the calculation by computing the electronic structure
for different values of external electrical fields without a sizeable
effect on the ${\rm d}I/{\rm d}V$ peak position. Finally, the modification
of the metallic surface is unlikely due to the very
weak physisorption interaction. These facts lead us to favor
H-bond mediated molecule-molecule
interaction, where the role of the surface may not be negligible due to the
important change of charge transfer that the orbital shift implies.

In summary, we found that PTCDA exhibits three distinct coexisting
adsorption phases on Au(788). The adsorption domains can be distinguished
by their geometric arrangement and by their electronic properties. By
performing DFT calculations the experimentally submolecular corrugation
is ascribed to the spatial distribution of the LUMO+1 density of states.
The shift of the LUMO+1 energy as seen in single-molecule spectroscopy
measurements in the distinct superstructures might be ascribed to the
different PTCDA packing densities and to the particular adsorption
geometries.

We are delighted to thank M.\ Sokolowski (University of Bonn, Germany)
for providing the PTCDA molecules and B.\ Hartke (University of Kiel,
Germany) for fruitful discussion. Financial support by the Deutsche
Forschungsgemeinschaft and the European Science Foundation is
gratefully acknowledged.

\end{document}